\documentclass[12pt]{article}

\usepackage{amsmath}
\usepackage{amssymb}
\usepackage{amsfonts}
\usepackage{latexsym}
\usepackage{color}

\catcode `\@=11 \@addtoreset{equation}{section}

\catcode `\@=12

\newcommand{\be}{\begin{equation}}
\newcommand{\en}{\end{equation}}
\newcommand{\bea}{\begin{eqnarray}}
\newcommand{\ena}{\end{eqnarray}}
\newcommand{\beano}{\begin{eqnarray*}}
\newcommand{\enano}{\end{eqnarray*}}
\newcommand{\bee}{\begin{enumerate}}
\newcommand{\ene}{\end{enumerate}}

\newcommand{\N}{\mathfrak N}

\newcommand{\mc}{\mathcal}

\newcommand{\F}{{\cal F}}

\newcommand{\1}{1 \!\! 1}

\newcommand{\Hil}{\mc H}
\newcommand{\kt}{\rangle}
\newcommand{\br}{\langle}

\catcode `\@=11 \@addtoreset{equation}{section}
\catcode `\@=12

\begin{document}

\thispagestyle{empty}

\vspace*{2cm}

\begin{center}
{\Large \bf Damping and Pseudo-fermions}   \vspace{2cm}\\

{\large F. Bagarello}\\
  Dieetcam,
Facolt\`a di Ingegneria,\\ Universit\`a di Palermo, I-90128  Palermo, Italy\\
e-mail: fabio.bagarello@unipa.it\\
home page: www.unipa.it$\backslash$fabio.bagarello

\end{center}

\vspace*{2cm}

\begin{abstract}
\noindent After a short abstract introduction on the time evolution driven by non self-adjoint hamiltonians, we show how the recently
introduced concept of {\em pseudo-fermion} can be used in the description of damping in finite dimensional quantum systems, and we compare the
results deduced adopting the Schr\"odinger and the Heisenberg representations.

\end{abstract}

\vspace{2cm}

%{\bf PACS Numbers}:  .......

\vfill

%\pagenumbering{roman}

\newpage

\section{Introduction}

In a series of papers, \cite{bagpb1}-\cite{bagrev}, we have considered two operators $a$ and $b$, with $b\neq a^\dagger$, acting on a Hilbert
space $\Hil$, and satisfying the  commutation rule $[a,b]=\1$. A nice functional structure has been deduced under suitable assumptions, and
some connections with physics, and in particular with what is usually called quasi-hermitian quantum mechanics and with the technique of
intertwining operators, have been established. Following Trifonov, \cite{tripb}, we have called {\em pseudo-bosons} (PB) the particle-like
excitations associated to this structure.  A similar analysis has also been carried out for what we have called {\em nonlinear pseudo-bosons}
(NLPB) in \cite{bagnlpb1}-\cite{bagzno2}, and most of the original results have been recovered also in this more general situation, where the
main ingredient is not the commutation rule between $a$ and $b$ but their raising and lowering properties, when applied to two fixed
biorthogonal bases. The analytical treatment of both PB and NLPB turns out to be particularly difficult in the case where {\em regularity} is
lost, that is, see below, when these bases are not Riesz bases. In this case, in fact, the intertwining operators appearing in the game (whose
square roots are metric operators in the sense of the literature on quasi-hermitian quantum mechanics, see \cite{ben,mosta,zno} and references
therein) turns out to be unbounded. For this reason, a large amount of {\em mathematical care} is required, and this makes the rigorous
treatment of the physical system rather complicated, \cite{bagzno2}.

More recently, \cite{bagPF}, we have introduced a similar extension for the canonical anticommutation relation (CAR), following an original
idea by Trifonov and his collaborators, \cite{tripf}, also briefly considered in \cite{most} and in \cite{bend2}. The idea is very similar to
the previous one: we consider again two operators $a$ and $b$, with $b\neq a^\dagger$, acting on a Hilbert space $\Hil$, and satisfying the
anticommutation rule $\{a,b\}=\1$, as well as $a^2=b^2=0$. Of course, if $b=a^\dagger$ we go back to ordinary CAR. The functional structure
that we deduce out of these operators is nice and friendly, since $\Hil$ is finite-dimensional. The related excitations are called {\em
pseudo-fermions}, (PF). One of the most important features of PF is that all the problems arising when dealing with PB, because of their (in
general) unbounded nature, are now absent.

In this paper we focus on a particular aspect of PF, i.e. on their possible use in the analysis of damped quantum systems. More in details: in
the description of damping in quantum optics certain finite dimensional, non self-adjoint, matrices are quite often used. These matrices are
usually called {\em effective hamiltonians}. Solving the related Schr\"odinger equation, one sees that the wave-function of the system goes to
zero for large $t$, at least if the parameters of the model are chosen properly. Already in \cite{bagPF} we have seen that one such a model,
originally proposed in \cite{benaryeh}, can be easily written in terms of PF. However, we have not explored the damping features of that
system, because that was not our major interest. Here, on the other hand, we focus our attention exactly on this aspect, and we work both in
the Schr\"odinger and in the Heisenberg representations, showing that analogous conclusions can be deduced. We also extend our analysis to a
higher-dimensional Hilbert space, which we will relate to a two-dimensional family of PF. For both these examples we deduce damping
\footnote{It might be worth stressing that we are not claiming here that PF are {\bf always} relevant to explain damping.}.

The paper is organized as follows: in the next section we briefly discuss some basic facts on PF and on the time evolution in the Heisenberg
representation, when this is driven by some non self-adjoint hamiltonian. This, we believe, is useful since not all the readers are familiar
with this kind of time-evolution, and surely less readers are familiar with pseudo-fermions. Section III is devoted to our first,
two-dimensional, example, while in Section IV we consider a four-dimensional, example. For both these examples we compare the well known {\em
Schr\"odinger point of view} with the modified Heisenberg evolution, and we show that, not surprisingly, the same conclusions can be deduced.
Particularly relevant for us is the fact that, after a somehow standard transformation of the effective hamiltonian, we will naturally driven
to introduce pseudo-fermions in the game. Section V contains an abstract generalization of the approach, while our conclusions are given in
Section VI.

\section{Pseudo-fermions and dynamics}

We begin this section by briefly reviewing the main definitions and results concerning PF in one dimension. The extension to higher dimensions
will be discussed later on. The starting point is a modification of the CAR $\{c,c^\dagger\}=c\,c^\dagger+c^\dagger\,c=\1$,
$\{c,c\}=\{c^\dagger,c^\dagger\}=0$, between two operators, $c$ and $c^\dagger$, acting on a two-dimensional Hilbert space $\Hil$. The CAR are
replaced here by the following rules: \be \{a,b\}=\1, \quad \{a,a\}=0,\quad \{b,b\}=0, \label{220}\en where the interesting situation is when
$b\neq a^\dagger$. These rules automatically imply that a non zero vector, $\varphi_0$, exists in $\Hil$ such that $a\,\varphi_0=0$, and that a
second non zero vector, $\Psi_0$, also exists in $\Hil$ such that $b^\dagger\,\Psi_0=0$, \cite{bagPF}.

Under these conditions it is possible to recover similar results as those for PB. In particular, we first introduce the following non zero
vectors \be \varphi_1:=b\varphi_0,\quad \Psi_1=a^\dagger \Psi_0, \label{221}\en as well as the non self-adjoint operators \be N=ba,\quad
\N=N^\dagger=a^\dagger b^\dagger. \label{222}\en We further introduce the self-adjoint operators $S_\varphi$ and $S_\Psi$ via their action on a
generic $f\in\Hil$: \be S_\varphi f=\sum_{n=0}^1\br\varphi_n,f\kt\,\varphi_n, \quad S_\Psi f=\sum_{n=0}^1\br\Psi_n,f\kt\,\Psi_n. \label{223}\en
Hence we get the following results,   whose proofs are straightforward and will not be given here:

\begin{enumerate}

\item \be a\varphi_1=\varphi_0,\quad b^\dagger\Psi_1=\Psi_0.\label{224}\en

\item \be N\varphi_n=n\varphi_n,\quad \N\Psi_n=n\Psi_n,\label{225}\en
for $n=0,1$.

\item If the normalizations of $\varphi_0$ and $\Psi_0$ are chosen in such a way that $\left<\varphi_0,\Psi_0\right>=1$,
then \be \left<\varphi_k,\Psi_n\right>=\delta_{k,n},\label{226}\en for $k,n=0,1$.

\item $S_\varphi$ and $S_\Psi$ are bounded, strictly positive, self-adjoint, and invertible. They satisfy
\be \|S_\varphi\|\leq\|\varphi_0\|^2+\|\varphi_1\|^2, \quad \|S_\Psi\|\leq\|\Psi_0\|^2+\|\Psi_1\|^2,\label{227}\en \be S_\varphi
\Psi_n=\varphi_n,\qquad S_\Psi \varphi_n=\Psi_n,\label{228}\en for $n=0,1$, as well as $S_\varphi=S_\Psi^{-1}$. Moreover, the following
intertwining relations \be S_\Psi N=\N S_\Psi,\qquad S_\varphi \N=N S_\varphi,\label{229}\en hold.

\end{enumerate}

The above formulas show that (i) $N$ and $\N$ behave as fermionic number operators, having eigenvalues 0 and 1; (ii) their related eigenvectors
are respectively the vectors of $\F_\varphi=\{\varphi_0,\varphi_1\}$ and $\F_\Psi=\{\Psi_0,\Psi_1\}$; (iii) $a$ and $b^\dagger$ are lowering
operators for $\F_\varphi$ and $\F_\Psi$ respectively; (iv) $b$ and $a^\dagger$ are rising operators for $\F_\varphi$ and $\F_\Psi$
respectively; (v) the two sets $\F_\varphi$ and $\F_\Psi$ are biorthonormal; (vi) the {\em well-behaved}\footnote{i.e. self-adjoint, bounded, invertible, with bounded inverse} operators $S_\varphi$ and
$S_\Psi$  maps $\F_\varphi$ in $\F_\Psi$ and viceversa; (vii) $S_\varphi$ and $S_\Psi$ intertwine between operators which are not self-adjoint,
in the very same way as they do for PB.

We refer to \cite{bagPF} for further remarks and consequences of these definitions. In particular, for instance, it is shown that $\F_\varphi$
and $\F_\Psi$ are automatically Riesz bases for $\Hil$, and the relations between fermions and PF are discussed.

\subsection{The algebraic dynamics}
\label{sectad}

In \cite{benaryeh} a non self-adjoint {\em effective} hamiltonian is considered in the description of the time evolution of the wave-function
of a two-dimensional quantum system. The resulting dynamics shows a decay of this wave-function. This is just a single reference where such an
approach is considered. Other references are, for instance, \cite{tripf}, \cite{graefe} and, for more abstract considerations,
\cite{bagznoproc}. The starting point is always the same: a Schr\"odinger equation \be i\dot\Psi(t)=H_{eff}\Psi(t),\label{ad1}\en where
$H_{eff}\neq H_{eff}^\dagger$. Actually, in \cite{benaryeh} and \cite{tripf}, $H_{eff}$ is assumed to be pseudo-hermitian. We will not consider
this extra assumption in the first part of this section. It is worth stressing here that formula (\ref{ad1}) should be considered  as an
assumption in itself, since in standard quantum mechanics the Schr\"odinger equation is usually (if not always) assumed to hold for a
self-adjoint hamiltonian. Of course, if $H_{eff}$ does not depend explicitly on time, the formal solution of (\ref{ad1}) is
$\Psi(t)=e^{-iH_{eff}t}\Psi(0).$ It is clear that, if $H_{eff}$ is a finite-dimensional matrix, this exponential can be defined using, for
instance, a norm convergent series:
$$
e^{-iH_{eff}t}:=\|.\|-\sum_{k=0}^\infty \frac{(-it)^k}{k!}\,H_{eff}^k.
$$
On the other hand, if $H_{eff}$ is an unbounded operator, defining $e^{-iH_{eff}t}$ is  quite a delicate task, since we cannot even use the
spectral theorem. We will not consider this aspect of the theory here, since we are concerned only with finite matrices. Solving (\ref{ad1})
means that we are adopting the {\em Schr\"odinger representation}. To move to the {\em Heisenberg representation}, we assume, as it is always
done in ordinary quantum mechanics, that the time evolution of the mean values of the observables do not depend on the representation chosen.
In other words, if $X$ is an observable of our physical system, calling $X_{eff}(t)$ its time evolution driven by $H_{eff}$, we require that
$$
\left<\Psi(t),X\Psi(t)\right>=\left<\Psi(0),X_{eff}(t)\Psi(0)\right>.
$$
Since this equality must hold for all possible choices of $\Psi(0)$, we deduce that \be X_{eff}(t):=e^{iH_{eff}^\dagger\,t}X\,e^{-iH_{eff}\,t}.
\label{ad2}\en This is the time evolution of $X$ in the Heisenberg representation. Notice that, if $H_{eff}=H_{eff}^\dagger=:H$, we go back to
the usual formula and to the unitary evolution, $X(t):=e^{iH\,t}X\,e^{-iH\,t}$. Notice also that, while $(XY)(t)=X(t)Y(t)$, $(XY)_{eff}(t)\neq
X_{eff}(t)Y_{eff}(t)$: the time evolution in (\ref{ad2}) is not an automorphism of the set of the observables of the system. Another
interesting feature of (\ref{ad2}) is that this time evolution is {\em stable under the adjoint}: $(X_{eff}(t))^\dagger=(X^\dagger)_{eff}(t)$.

Also, it is clear that the Heisenberg equation of motion for $X_{eff}(t)$ should be modified. For that, let us define the following {\em
effective commutator} between two generic operators $A$ and $B$: \be [A,B]_{eff}:=A\,B-B^\dagger\, A. \label{ad3}\en The effective commutator
satisfies, for example, the following equality: given three generic operators $A$, $B$ and $C$, then
$$
[AB,C]_{eff}=A[B,C]_{eff}+[A,C]_{eff}\,B+A(C^\dagger-C)B,
$$
which reduces to the standard formula when $C=C^\dagger$, but not in general. Also, not surprisingly, the Jacobi identity ceases to be true for
$[.,.]_{eff}$. Still we have $[A,B]_{eff}^\dagger=-[A^\dagger,B]_{eff}$.

The  differential equation for $X_{eff}(t)$ is the following: \be
i\frac{d}{dt}\,X_{eff}(t)=\left[X_{eff}(t),H_{eff}\right]_{eff}=e^{iH_{eff}^\dagger\,t}[X,H_{eff}]_{eff}e^{-iH_{eff}\,t}, \label{ad4}\en which
returns the standard equation if $H_{eff}=H_{eff}^\dagger$. This formula shows, among other things, that an integral of motion, $Z$, is not an
operator commuting with $H_{eff}$. What $Z$ has to satisfy is clearly the following requirement: $[Z,H_{eff}]_{eff}=0$. In fact, if this is
true, then $\dot Z_{eff}(t)=0$.

In \cite{graefe} the authors have considered a slightly different point of view, splitting $H_{eff}$ into two parts, both self-adjoint:
$H_{eff}=\frac{1}{2}\left(H_{eff}+H_{eff}^\dagger\right)-i\,\frac{1}{2i}\left(H_{eff}^\dagger-H_{eff}\right)$. In this way, rather than
introducing an effective commutator, they deduce a differential equation for $X_{eff}(t)$ in which both a commutator and an anti-commutator
appear.

Something more can be said if $H_{eff}$ is crypto-hermitian, see \cite{bagzno1}, i.e. if a positive self-adjoint, time-independent, operator
$\Theta$ exists, bounded with bounded inverse, such that $H_{eff}=\Theta^{-1}H_{eff}^\dagger\Theta$. In this case we get the following:
$$
\Theta^{-1}X_{eff}(t)=e^{iH_{eff}t}\left(\Theta^{-1}X\right)e^{-iH_{eff}t},
$$
and the differential equation is $$\frac{d}{dt}\left(\Theta^{-1}X_{eff}(t)\right)= i\,e^{iH_{eff}t}[H_{eff},\Theta^{-1}X]e^{-iH_{eff}t}.
$$
Let us remark that here, first of all, the commutator is the standard one. Also, at a first sight the time evolution for $\Theta^{-1}X$ might
appear as an automorphism. Nevertheless, it is not hard to check that this is not true, exactly because of the presence of $\Theta^{-1}$. In fact,
$\Theta^{-1}X_{eff}(t)$ is, in general, different from $(\Theta^{-1}X)_{eff}(t)$. A simple case when they coincide is when $H_{eff}$ is
self-adjoint. The essential reason why formulas above look interesting is because only $H_{eff}$ appears, and not $H_{eff}^\dagger$. In
particular, the differential equation shows that $X_{eff}(t)$ is constant when $H_{eff}$ commutes not with $X$ but with $\Theta^{-1}X$. This is the
formula which replaces, in this particular case, the one deduced above, $[X,H_{eff}]_{eff}=0$.

Because of the properties of $\Theta$, we know that $\Theta^{\pm 1/2}$ can be defined, and $h:=\Theta^{1/2}H_{eff}\Theta^{-1/2}$ is a
self-adjoint operator: $h=h^\dagger$. Let us now introduce the following automorphism of $B(\Hil)$, the set of the bounded operators on $\Hil$:
$j_\Theta(x):=\Theta^{-1/2}x\,\Theta^{1/2}$. This map is invertible, and it is clear that $h=j_\Theta^{-1}(H_{eff})$. If we further define the
following {\em standard Heisenberg evolution}, $\alpha^t(X):=e^{iht}Xe^{-iht}$, $X\in B(\Hil)$, which has all the usual properties of the time
evolution for systems driven by self-adjoint hamiltonians, we deduce that \be X_{eff}(t)=j_\Theta^{-1}\left(\alpha^t(j_\Theta(X))\right),
\label{ad5}\en which has a first interesting consequence: this effective time evolution coincides with $\alpha^t$ if this latter commutes with
$j_\Theta$. Moreover, the interpretation is the following: to find $X_{eff}(t)$, we have first to map $X$  into $j_\Theta(X)$. Then we let this
operator evolve using $h$, and finally we transform back this result, using $j_\Theta^{-1}$; therefore, the effective time evolution appears
to be the composition of these three maps.

\section{An example from the literature}

The first example is taken by \cite{tripf,benaryeh}, and its relation with PF was already partly considered in \cite{bagPF}. The starting point
is the Schr\"odinger equation \be i\dot \Psi(t)=H_{eff}\Psi(t), \quad\mbox{with}\quad  H_{eff}=\frac{1}{2}\left(
                                                       \begin{array}{cc}
                                                         -i\gamma_a & v \\
                                                         \overline{v} & -i\gamma_b \\
                                                       \end{array}
                                                     \right),
\label{30}\en where $\gamma_a, \gamma_b>0$ and $v\in{\Bbb C}$.

\subsection{Schr\"odinger representation}

With a simple change of variable $\Phi(t)=e^{\Gamma t}\Psi(t)$, $\Gamma=\frac{1}{2}(\gamma_a+\gamma_b)$, we deduce that $i\dot
\Phi(t)=H\Phi(t)$, where
$$
H=i\Gamma\1_2+H_{eff}=\left(
                                                       \begin{array}{cc}
                                                         -i\gamma & v \\
                                                         \overline{v} & i\gamma \\
                                                       \end{array}
                                                     \right),\qquad
\Phi(t)=\left(
          \begin{array}{c}
            \Phi_0(t) \\
            \Phi_1(t) \\
          \end{array}
        \right).
$$
Here $\1_2$ is the two-by-two identity matrix, and $\gamma=\frac{1}{2}(\gamma_a-\gamma_b)$. The differential equations for $\Phi_0(t)$ and
$\Phi_1(t)$ are easily deduced. Let us introduce $\Omega:=|v|^2-\gamma^2$. Then \be \left\{
\begin{array}{ll}
                \ddot \Phi_0(t)=-\Omega\,\Phi_0(t),       \\
  \ddot \Phi_1(t)=-\Omega\,\Phi_1(t).
\end{array}
\right. \label{31}\en

If $\Omega=0$ then the functions $\Phi_0(t)$ and $\Phi_1(t)$ are linear in $t$, so that
$$
\Psi(t)=e^{-\Gamma t}\left(
          \begin{array}{c}
            \Phi_0(t) \\
            \Phi_1(t) \\
          \end{array}
        \right)=\left(
          \begin{array}{c}
            e^{-(\gamma_a+\gamma_b)\frac{t}{2}}(A_0+B_0\,t) \\
            e^{-(\gamma_a+\gamma_b)\frac{t}{2}}(A_1+B_1\,t) \\
          \end{array}
        \right),
$$
where $A_0, A_1, B_0$ and $B_1$ are fixed by the initial conditions. In particular, if $\Psi(0)=\left(
          \begin{array}{c}
            \varphi_0 \\
            \varphi_1 \\
          \end{array}
        \right)$, we find that $A_0=\varphi_0$, $A_1=\varphi_1$, $B_0=-\gamma\,\varphi_0-i\,v\,\varphi_1$ and $B_1=\gamma\,\varphi_1-i\,\overline{v}
        \,\varphi_0$.

 The norm of $\Psi(t)$ is simply the norm in $\Hil={\Bbb C}^2$.
Therefore $$\|\Psi(t)\|^2=e^{-t(\gamma_a+\gamma_b)}\left(|A_0+B_0\,t|^2+|A_1+B_1\,t|^2\right),$$ and this squared norm goes to zero when
$t\rightarrow\infty$.

Let us consider now the case in which $\Omega>0$. In this case the solution can be written as
$$
\Psi(t)=e^{-(\gamma_a+\gamma_b)\frac{t}{2}}\,\left(
          \begin{array}{c}
            A_0\cos(\sqrt{\Omega}\,t)+B_0\sin(\sqrt{\Omega}\,t) \\
            A_1\cos(\sqrt{\Omega}\,t)+B_1\sin(\sqrt{\Omega}\,t) \\
          \end{array}
        \right),
$$
where, again, $A_0, A_1, B_0$ and $B_1$ are fixed by the initial conditions: $A_0=\varphi_0$, $A_1=\varphi_1$,
$B_0=\frac{1}{\sqrt{\Omega}}\left(-\gamma\,\varphi_0-i\,v\,\varphi_1\right)$ and
$B_1=\frac{1}{\sqrt{\Omega}}\left(\gamma\,\varphi_1-i\,\overline{v}
        \,\varphi_0\right)$.
 In this case $\|\Psi(t)\|^2$ is the product of a decaying exponential and an oscillating function. More explicitly,
$$
\|\Psi(t)\|^2=e^{-t(\gamma_a+\gamma_b)}\left(|A_0\cos(\sqrt{\Omega}\,t)+B_0\sin(\sqrt{\Omega}\,t)|^2+
|A_1\cos(\sqrt{\Omega}\,t)+B_1\sin(\sqrt{\Omega}\,t)|^2\right),
$$
which decays to zero independently of the particular values of $\varphi_0$ and $\varphi_1$.

Let us finally consider  the case in which $\Omega<0$. In this case the solution can be written as
$$
\Psi(t)=e^{-(\gamma_a+\gamma_b)\frac{t}{2}}\,\left(
          \begin{array}{c}
            A_0\exp(\sqrt{|\Omega|}\,t)+B_0\exp(-\sqrt{|\Omega|}\,t) \\
            A_1\exp(\sqrt{|\Omega|}\,t)+B_1\exp(-\sqrt{|\Omega|}\,t) \\
          \end{array}
        \right).
$$
Here, $A_0, A_1, B_0$ and $B_1$ are related to the initial conditions as follows:
$$
A_0=\frac{1}{2}\left(\varphi_0+\frac{1}{\sqrt{|\Omega|}}\left(-\gamma\,\varphi_0-i\,v\,\varphi_1\right)\right),\quad
B_0=\frac{1}{2}\left(\varphi_0+\frac{1}{\sqrt{|\Omega|}}\left(\gamma\,\varphi_0+i\,v\,\varphi_1\right)\right),
$$
$$
A_1=\frac{1}{2}\left(\varphi_1+\frac{1}{\sqrt{|\Omega|}}\left(\gamma\,\varphi_1-i\,\overline{v}\,\varphi_0\right)\right),\quad
B_1=\frac{1}{2}\left(\varphi_1+\frac{1}{\sqrt{|\Omega|}}\left(-\gamma\,\varphi_1+i\,\overline{v}\,\varphi_0\right)\right).
$$
In this case it is not completely clear that $\|\Psi(t)\|$ goes to zero for increasing $t$. However, to get such a conclusion, it is enough to
consider the asymptotic behavior of $e^{-(\gamma_a+\gamma_b)t}\,e^{2\sqrt{|\Omega|}t}$, which is the only {\em dangerous} contribution
appearing in $\|\Psi(t)\|^2$. The conclusion is that $\|\Psi(t)\|\rightarrow0$, when $t\rightarrow\infty$,  if
$2\sqrt{|\Omega|}<\gamma_a+\gamma_b$, which is always satisfied.

The outcome of this analysis is, therefore, the following: {\em the wave-function $\Psi(t)$ solving the Schr\"odinger equation (\ref{30}) goes
to zero independently of the value of $\Omega$.}

\subsection{Heisenberg representation}

It is easy to see that $\Omega$ is related to the eigenvalues of $H$, which are $\lambda_\pm:=\pm\sqrt{|v|^2-\gamma^2}$. Hence
$\lambda_+=\lambda_-=0$ if $\Omega=0$, while $\lambda_\pm:=\pm\sqrt{\Omega}$ when $\Omega>0$. Finally, $\lambda_\pm:=\pm\,i\,\sqrt{|\Omega|}$
when $\Omega<0$. In all these cases the eigenstates of $H$ can be written as
$$
\eta_+=\left(
          \begin{array}{c}
            \frac{1}{\overline{v}}\left(-i\gamma+\sqrt{\Omega}\right) \\
            1 \\
          \end{array}
        \right),\qquad
\eta_-=\left(
          \begin{array}{c}
            -\,\frac{1}{\overline{v}}\left(i\gamma+\sqrt{\Omega}\right) \\
            1 \\
          \end{array}
        \right).
$$
In fact, $H\eta_\pm=\lambda_\pm\eta_\pm$. Moreover, in general $\eta_+$ and $\eta_-$ are not orthogonal:
$\left<\eta_+,\eta_-\right>=\frac{2\gamma}{|v|^2}\left(\gamma-i\sqrt{\Omega}\right)$, which is zero only if $\gamma=0$ ($H=H^\dagger$) or if
$\gamma=i\sqrt{\Omega}$ ($H=-H^\dagger$). Due to the relation between $H$ and $H_{eff}$ we also find that \be
H_{eff}\,\eta_\pm=E_\pm\,\eta_\pm, \qquad E_\pm=-\frac{i}{2}(\gamma_a+\gamma_b)\pm\sqrt{\Omega}. \label{32}\en In \cite{bagPF} we have
analyzed, in connection with PF, the following non self-adjoint operator,
$$ \tilde H=\frac{1}{2}\left(
                                                       \begin{array}{cc}
                                                         -i\delta & \overline{\omega} \\
                                                         \omega & i\delta \\
                                                       \end{array}
                                                     \right),
$$ which coincides with our $H$ simply taking $\delta=2\gamma$ and $\omega=2\overline{v}$. As in \cite{bagPF}, it is therefore possible to introduce
two operators $a$ and $b$, such that $\{a,b\}=\1$, $a^2=b^2=0$, and $ H=\Omega\,\left(b\,a-\frac{1}{2}\,\1\right)$. Introducing also the number
operators $N=b\,a$ and $\N=N^\dagger$, and assuming, just to fix the ideas, that $\Omega>0$, we can write
$H=\Omega\,\left(N-\frac{1}{2}\,\1\right)$ and $H^\dagger=\Omega\,\left(\N-\frac{1}{2}\,\1\right)$. Now, we want to recover here the same
damping we have found working in Schr\"odinger representation. For that, it is natural to consider the time evolution of the number operators
$N$ and $\N$, which should be computed as in (\ref{ad2}):
$$
N_{eff}(t)=e^{iH_{eff}^\dagger\,t}N\,e^{-iH_{eff}\,t}, \qquad \N_{eff}(t)=e^{iH_{eff}^\dagger\,t}\N\,e^{-iH_{eff}\,t}=(N_{eff}(t))^\dagger.
$$
Due to the relation between $H_{eff}$ and $H$, $H_{eff}=H-i\Gamma\1$, and to the above expressions for $H$ and $H^\dagger$, we find that, for
each observable $X$, \be X_{eff}(t)=e^{-2\Gamma\,t}e^{i\Omega\N t}\,X\, e^{-i\Omega N t}.\label{33}\en Therefore, recalling that the
anticommutation rules imply that $N^2=N$, $\N^2=\N$, and so on, we find that
$$
N_{eff}(t)=e^{-2\Gamma\,t}\left(Ne^{-i\Omega t}+\N\,N(1-e^{-i\Omega t})\right),
$$
while $\N_{eff}(t)$ is simply its adjoint. Then, if we estimate the norm of $N_{eff}(t)$, it is trivial to deduce that $\|N_{eff}(t)\|\leq
3e^{-2\Gamma t}$, which goes to zero when $t$ diverges. Hence, as expected, we recover damping also in  Heisenberg picture, with no need of
introducing extra assumptions on the parameters of the system. More connections with PF are discussed in \cite{bagPF}.

\section{A two-dimensional example}

We are going now to repeat the same analysis considering the following four-by-four effective hamiltonian:
$$
H_{eff}=\frac{i}{\alpha-\beta}\left(
                                \begin{array}{cccc}
                                  0 & -\alpha\beta(\omega_1-\omega_2) & 0 & 0 \\
                                  \omega_1-\omega_2 & -(\alpha+\beta)(\omega_1-\omega_2) & 0 & 0 \\
                                  -\omega_2 & \alpha\omega_2 & \beta\omega_2-\alpha\omega_1 & 0 \\
                                  -\beta\omega_2 & \beta^2\omega_2 & 0 & \alpha\omega_2-\beta\omega_1 \\
                                \end{array}
                              \right),
$$
where $\alpha\neq\beta$ and $\omega_1>\omega_2>0$. It is evident that $H_{eff}\neq H_{eff}^\dagger$. We will show that, also for this
hamiltonian, a decay can be deduced.

\subsection{Schr\"odinger representation}

As before, we want to solve first the Schr\"odinger equation $i\dot \Psi(t)=H_{eff}\Psi(t)$ and check whether damping is deduced, and in which
conditions. Even if $H_{eff}$ has no direct physical interpretation, still it is interesting, since, with a simple transformation, we can
transform the original Schr\"odinger equation into a new one, with a traceless hamiltonian $H$, which is also not self-adjoint. This is very
close to what was done, for instance, in \cite{tripf,bagPF,benaryeh}, and in the previous section.  For that it is enough to proceed as in the
previous example, defining $\Phi(t)=e^{\Gamma t}\Psi(t)$, $H=i\Gamma\1_4+H_{eff}$, and
$\Gamma=\frac{\alpha+\beta}{2(\alpha-\beta)}\,(\omega_1-\omega_2)$, where $\1_4$ is the identity matrix in $\Hil={\Bbb C}^4$. Then we get
$$
i\,\dot \Phi(t)=H\Phi(t),$$ where \be {\small H=\frac{i}{\alpha-\beta}\left(
                                \begin{array}{cccc}
                                  \frac{1}{2}\,(\alpha+\beta)(\omega_1-\omega_2) & -\alpha\beta(\omega_1-\omega_2) & 0 & 0 \\
                                  \omega_1-\omega_2 & -\frac{1}{2}\,(\alpha+\beta)(\omega_1-\omega_2) & 0 & 0 \\
                                  -\omega_2 & \alpha\omega_2 & -\frac{1}{2}\,(\alpha-\beta)(\omega_1+\omega_2) & 0 \\
                                  -\beta\omega_2 & \beta^2\omega_2 & 0 & \frac{1}{2}\,(\alpha-\beta)(\omega_1+\omega_2) \\
                                \end{array}
                              \right),}
\label{34}\en and $$ \Phi(t)=\left(
                               \begin{array}{c}
                                 \Phi_0(t) \\
                                 \Phi_1(t) \\
                                 \Phi_2(t) \\
                                 \Phi_3(t) \\
                               \end{array}
                             \right).
$$
As in our first example, the matrix $H$ is traceless. The related Schr\"odinger equation of motion produces, first of all, $\ddot \Phi_0(t)=\Omega\,\Phi_0(t)$, where $\Omega:=\frac{1}{2}|\omega_1-\omega_2|$. Then
$\Phi_0(t)=\tilde A_0\,e^{\Omega\,t}+\tilde B_0\,e^{-\Omega\,t}$, with $\tilde A_0$ and $\tilde B_0$ fixed by the initial conditions.
The second component of $\Phi(t)$, $\Phi_1(t)$, can be deduced from $\Phi_0(t)$ since
$$
\Phi_1(t)=\frac{1}{\alpha\beta(\omega_1-\omega_2)}\left(\frac{1}{2}\,(\alpha+\beta)(\omega_1-\omega_2)\Phi_0(t)-(\alpha-\beta)\dot\Phi_0(t)\right).
$$
Moreover
\be \left\{
\begin{array}{ll}
                \Phi_2(t)=A_2\,e^{-(\omega_1+\omega_2)t/2}+\tilde A_2 \,e^{\Omega t}+\tilde B_2 \,e^{-\Omega t},       \\
   \Phi_3(t)=A_3\,e^{(\omega_1+\omega_2)t/2}+\tilde A_3 \,e^{\Omega t}+\tilde B_3 \,e^{-\Omega t}.
\end{array}
\right. \label{35}\en
As for the asymptotic behavior of $\Psi(t)$, we recall that $\omega_1>\omega_2>0$. Of course, to conclude that $\Psi(t)=e^{-\Gamma t}\Phi(t)\rightarrow0$
when
$t\rightarrow\infty$, it is enough to consider the {\em worse} contribution, i.e. the one coming from $\Phi_3(t)$: in fact, if this goes to zero,
then it is clear that all the other components go to zero as well, so that $\|\Psi(t)\|\rightarrow 0$ when $t\rightarrow\infty$. We have, for $t\gg1$,
$$
\Psi_3(t)\simeq \exp\left\{\left(\frac{1}{2}(\omega_1+\omega_2)-\Gamma\right)t\right\}=
\exp\left\{\frac{t}{\alpha-\beta}\,\left(\alpha\omega_2-\beta\omega_1\right)\right\}.
$$
Recalling that $\alpha>\beta$, $\Psi_3(t)\rightarrow0$ if and only if $\alpha\omega_2-\beta\omega_1<0$, i.e. if
\be\frac{\alpha}{\beta}<\frac{\omega_1}{\omega_2}.\label{36}\en
In other words, when condition (\ref{36}) is satisfied, not only $\Psi_3(t)$, but $\Psi(t)$ itself decreases to zero for $t$ diverging.

\subsection{Heisenberg representation}

Let us now consider the following operators
$$
a_1:=\frac{1}{\alpha}\left(
                       \begin{array}{cccc}
                         \frac{-\beta^2}{\alpha-\beta} & \frac{\beta^3}{\alpha-\beta} & 0 & \beta \\
                         \frac{-\beta}{\alpha-\beta} & \frac{\beta^2}{\alpha-\beta} & 0 & 1 \\
                         \frac{\alpha^2-\beta}{\alpha-\beta} & \frac{\beta(-\alpha^2+\beta)}{\alpha-\beta} & 0 & 1 \\
                         0 & 0 & 0 & 0 \\
                       \end{array}
                     \right),\quad
                     b_1:=\left(
                       \begin{array}{cccc}
                         \frac{1}{\alpha-\beta} & \frac{-\alpha}{\alpha-\beta} & 1 & 0 \\
                         \frac{1}{\alpha(\alpha-\beta)} & \frac{-1}{\alpha-\beta} & \frac{1}{\alpha} & 0 \\
                         0 & 0 & 0 & 0 \\
                         \frac{-\alpha^2+\beta}{\alpha(\alpha-\beta)} & \frac{\alpha^2-\beta}{\alpha-\beta} & \frac{\beta}{\alpha} & 0 \\
                       \end{array}
                     \right),
$$
$$
a_2:=\left(
                       \begin{array}{cccc}
                         \frac{\beta}{\alpha-\beta} & \frac{-\beta^2}{\alpha-\beta} & 0 & -1 \\
                         \frac{\beta}{\alpha(\alpha-\beta)} & \frac{-\beta^2}{\alpha(\alpha-\beta)} & 0 & -\frac{1}{\alpha} \\
                         \frac{-\alpha}{\alpha-\beta} & \frac{\alpha^2}{\alpha-\beta} & 0 & 0 \\
                         \frac{\beta^2}{\alpha(\alpha-\beta)} & \frac{-\beta^3}{\alpha(\alpha-\beta)} & 0 & -\frac{\beta}{\alpha} \\
                       \end{array}
                     \right),\quad
                     b_2:=\left(
                       \begin{array}{cccc}
                         \frac{\beta}{\alpha(\alpha-\beta)} & \frac{-\beta}{\alpha-\beta} & \frac{\beta}{\alpha} & 0 \\
                         \frac{1}{\alpha(\alpha-\beta)} & \frac{-1}{\alpha-\beta} & \frac{1}{\alpha} & 0 \\
                         \frac{1}{\alpha(\alpha-\beta)} & \frac{-1}{\alpha-\beta} & \frac{1}{\alpha} & 0 \\
                         \frac{-\alpha}{\alpha-\beta} & \frac{\alpha\beta}{\alpha-\beta}  & 0 & 0 \\
                       \end{array}
                     \right).
$$
They satisfy the following anticommutation rules: $\{a_j,b_k\}=\delta_{j,k}\1_4$, with $\{a_j,a_k\}=\{b_j,b_k\}=0$, $j,k=1,2$\footnote{Other
examples of $4 \times 4$ matrices satisfying these rules can be found in \cite{trinew}}. Therefore they are two-dimensional pseudo-fermions. $H$ can
be written in terms of these operators as
$$
H=i\left(\omega_1\,b_1\,a_1+\omega_2\,b_2\,a_2-\frac{\omega_1+\omega_2}{2}\,\1_4\right).
$$
As in the previous example, the time evolution of a given operator $X$, can be written in a way which extends (\ref{33}):
$$
X_{eff}(t)=e^{iH_{eff}^\dagger t}\,X\,e^{-iH_{eff} t}=e^{-(2\Gamma+\omega_1+\omega_2)t}e^{(\omega_1\N_1+\omega_2\N_2)t}\,X\,
e^{(\omega_1N_1+\omega_2N_2)t},
$$
where $N_j=b_ja_j$ and $\N_j=N_j^\dagger$, $j=1,2$. Recalling that $[N_1,N_2]=0$ and that $N_j^2=N_j$, we get
$$
N_{1,eff}(t)=e^{-(2\Gamma+\omega_2)t}\left(N_1+\N_1\,N_1(e^{\omega_1t}-1)\right)\left(\1_4+(N_2+\N_2)(e^{\omega_2t}-1)+
\N_2\,N_2(e^{\omega_2t}-1)\right),
$$
and a similar expression can be deduced for $N_{2,eff}(t)$. Moreover, $\N_{j,eff}(t)$ is simply the adjoint of $N_{j,eff}(t)$. Using
$\|N_j\|\leq1$ and  $\|\N_j\|\leq1$, $j=1,2$, we get
$$
\|N_{1,eff}(t)\|\leq \exp\{-(2\Gamma-\omega_1-\omega_2)t\},
$$
which goes to zero, when $t\rightarrow\infty$, if and only if $2\Gamma>\omega_1+\omega_2$, i.e., if condition (\ref{36}) is satisfied: we
recover exactly (and not surprisingly) the same conclusion as in the Schr\"odinger representation.

\vspace{3mm}

{\bf Remark:--} It might be interesting to notice that the matrices $a_j$ and $b_j$ above can be related to the following, standard,
two-dimensional fermion annihilation  operators,
$$
A_1=\left(
      \begin{array}{cccc}
        0 & 1 & 0 & 0 \\
        0 & 0 & 0 & 0 \\
        0 & 0 & 0 & 1 \\
        0 & 0 & 0 & 0 \\
      \end{array}
    \right),\qquad A_2=\left(
      \begin{array}{cccc}
        0 & 0 & 1 & 0 \\
        0 & 0 & 0 & -1 \\
        0 & 0 & 0 & 0 \\
        0 & 0 & 0 & 0 \\
      \end{array}
    \right),
$$
as in \cite{bagPF}: $a_j=TA_jT^{-1}$, $b_j=TA_j^\dagger T^{-1}$, $j=1,2$, where $T$ is the following non-singular matrix:
$$
T=\left(
      \begin{array}{cccc}
        0 & \alpha & \beta & 0 \\
        0 & 1 & 1 & 0 \\
        \alpha & 0 & 1 & 0 \\
        0 & \beta & 0 & \alpha \\
      \end{array}
    \right).
$$
Here $\alpha\neq0$ and $\alpha\neq\beta$. Therefore, this example can be considered as an explicit realization of the equivalence theorem
stated in \cite{bagPF}, concerning the relations between fermions and pseudo-fermions.

\section{An abstract generalization}

We will now describe how the examples considered in the previous sections can be further extended.

Let us consider two sets of pseudo-fermionic operators, $a_j$ and $b_j$, with $b_j\neq a_j^\dagger$, $j=1,2,\ldots,N$, satisfying
$$
\{a_j,b_k\}=\delta_{j,k}\,\1,\qquad \{a_j,a_k\}=\{b_j,b_k\}=0,
$$
for all $j$ and $k$, and let $\varphi_{0,0,\ldots,0}$ and $\Psi_{0,0,\ldots,0}$, with
$\left<\varphi_{0,0,\ldots,0},\Psi_{0,0,\ldots,0}\right>=1$, be the vacua of the $a_j$'s and $b_j^\dagger$'s respectively:
$$a_j\varphi_{0,0,\ldots,0}=b_j^\dagger\Psi_{0,0,\ldots,0}=0,$$
for all $j$. Then, extending the procedure of Section II, the sets $\F_\varphi$ and $\F_\Psi$ of the functions
$$
\varphi_{\underline n}:=\varphi_{n_1,n_2,\ldots,n_N}=b_1^{n_1}b_2^{n_2}\cdots b_N^{n_N}\varphi_{0,0,\ldots,0},
$$
and
$$
\Psi_{\underline n}:=\Psi_{n_1,n_2,\ldots,n_N}=(a_1^\dagger)^{n_1}(a_2^\dagger)^{n_2}\cdots (a_N^\dagger)^{n_N}\Psi_{0,0,\ldots,0},
$$
$n_j=0,1$, are biorthonormal bases of the $2^N$-dimensional Hilbert space $\Hil_N$.

Let now $i\dot\Psi(t)=H_{eff}\Psi(t)$, with $H_{eff}\neq H_{eff}^\dagger$, be our {\em original} Schr\"odinger equation. As in Sections III and
IV, we introduce a new wave function, $\Phi(t)=e^{\Gamma t}\Psi(t)$, where $\Gamma$ is a real constant, to be fixed. The Schr\"odinger equation
for $\Phi(t)$ is
$$
i\dot\Phi(t)=H_N\Phi(t),\qquad H_N:=H_{eff}+i\Gamma\1_N,
$$
where $\1_N$ is the identity operator on $\Hil_N$. The examples discussed before show that, in some interesting situations, the hamiltonian
$H_N$ has the following general form:
$$
H_N=\sum_{j=1}^N\,\Omega_j\,N_j-\frac{1}{2}\sum_{j=1}^N\,\Omega_j\1_N,
$$
where $N_j=b_j\,a_j$ is the $j-th$ pseudo-fermionic number operator, while the $\Omega_j$'s are, in general, complex quantities. Let us now
define the following quantity: \be T_N:=\sum_{{\underline k},{\underline n}}\left<\Psi_{{\underline n}},H_N\varphi_{{\underline
k}}\right>=\sum_{{\underline k}}\left<\Psi_{\underline k},H_N\varphi_{\underline k}\right>, \label{51}\en where the last equality is a
consequence of the orthonormality of  $\F_\varphi$ and $\F_\Psi$, and of the fact that $\varphi_{\underline k}$ is eigenstate of $H_N$.   $T_N$
appears a natural extension of the trace of an operator to the present settings, where a single orthonormal basis is replaced by two
biorthonormal bases. From (\ref{51}) it is now easy to check that $T_N=0$ for all $N=1,2,\ldots$.

In the (extended) Heisenberg representation, see Section \ref{sectad}, the time evolution $X_{eff}(t)=e^{i H_{eff}^\dagger t}\,X\,e^{-i H_{eff}
t}$ of a given operator $X$, assume the following form: \be
X_{eff}(t)=e^{-t(2\Gamma+\Im(\Omega_1)+\cdots+\Im(\Omega_N))}\prod_{j=1}^N\,e^{it\,\overline{\Omega_j}\,\N_j}\,X\,\prod_{l=1}^N\, e^{it
\Omega_l\,N_l}. \label{52}\en Focusing now to the time behavior of the number operator $N_k$, and recalling that $N_j^2=N_j$ and $\N_j^2=\N_j$,
for all $j$, we deduce that \be
(N_k)_{eff}(t)=e^{-t(2\Gamma+\Im(\Omega_1)+\cdots+\Im(\Omega_N))}\prod_{j=1}^N\,\left(\1_N+\N_j\left(e^{it\,\overline{\Omega_j}}-1\right)\right)N_k
\prod_{l=1}^N\,\left(\1_N+N_l\left(e^{-it\,\Omega_l}-1\right)\right) \label{53}\en Damping can be found by estimating the norm of the various
$(N_k)_{eff}(t)$. For that, we first recall that $\|N_j\|=\|\N_j\|\leq1$, for all $j$. The simplest situation is when the complex frequencies
are all real: in fact, if $\Im(\Omega_j)=0$ for all $j$, we easily conclude that
$$
\|(N_k)_{eff}(t)\|\leq 3^{2N}\,e^{-2\Gamma t},
$$
which goes to zero when $t$ diverges for all possible values of positive $\Gamma$. Hence, in this case, damping is deduced. It is probably more
interesting to consider the situation in which not all the $\Omega_j$'s are strictly real. In particular, we consider here the opposite
situation, i.e. the one in which $\Im(\Omega_j)\neq0$ for all $j$, since this is the most {\em dangerous} case. Now, the estimate above should
be replaced by the following:
$$
\|(N_k)_{eff}(t)\|\leq e^{-(2\Gamma+\Im(\Omega_1)+\cdots+\Im(\Omega_N)) t}\prod_{j=1}^N\left(2+e^{t\Im(\Omega_j)}\right)^2,
$$
which, clearly, extends the previous one. In this case, damping is recovered if \be \Gamma>\frac{1}{2}\sum_{j=1}^N|\Im(\Omega_j)|,
\label{add1}\en while a similar (and simpler) condition is recovered if only some of the $\Omega_j$'s are not purely real. Notice that the
inequality (\ref{add1}) becomes quite simple, $\Gamma>0$, when $\Im(\Omega_j)=0$, for all $j$. In this case, we recover our previous
conclusion.

If, in analogy with $T_N$, we introduce the {\em generalized} trace also for $H_{eff}$, recalling that $T_N=0$ we find that
$$
T_N^{eff}:=\sum_{\underline k}\left<\Psi_{\underline k},H_{eff}\varphi_{\underline k}\right>=-i2^N\Gamma.
$$
Therefore, in order for $H_{eff}$ to produce damping, the following inequality should be satisfied by the effective hamiltonian $H_{eff}$:
$$
i\,T_N^{eff}>2^{N-1}\sum_{j=1}^N|\Im(\Omega_j)|.
$$

 \vspace{4mm}

We end this section with some remarks on the pseudo-fermionic structure related to this abstract system. The starting point is the pair of
operators $S_\varphi$ and $S_\Psi$ which extends those of (\ref{223}). They satisfy a multi-dimensional version of (\ref{229}),
$$
N_j\,S_\varphi=S_\varphi\,\N_j,\qquad \N_j\,S_\Psi=S_\Psi\,N_j,
$$
for all $j$. Moreover, since the square root of both $S_\varphi$ and $S_\Psi$ can be defined, we can also introduce the self-adjoint operators
$n_j=S_\Psi^{1/2}\,N_j\,S_\varphi^{1/2}=n_j^\dagger$. Therefore, if the $\Omega_j$ are real, we deduce that $S_\Psi\,H_N=H_N^\dagger\,S_\Psi$
and that \be h:=S_\Psi^{1/2}\,H_N\,S_\varphi^{1/2}=\sum_{j=1}^N\,\Omega_j\,n_j-\frac{1}{2}\sum_{j=1}^N\,\Omega_j\1_N \label{54}\en is a
self-adjoint operator.

It is well known that the operator $S_\Psi^{1/2}$, as well as its inverse $S_\varphi^{1/2}$, can be  used to define a different scalar product
in $\Hil_N$. Then the conclusion is the following: a damping effect in $\Hil_N$, endowed with its {\em natural} scalar product $<.,.>$, can be
obtained using certain non self-adjoint effective hamiltonians which, when considered again in $\Hil_N$, but endowed with a different scalar
product, are self-adjoint. In fact, let us define a new scalar product $\left<.,.\right>_\Psi$ on $\Hil_N$ as follows:
$$
\left<f,g\right>_\Psi:=\left<S_\Psi^{1/2}\,f,S_\Psi^{1/2}\,g\right>,
$$
for all $f,g\in\Hil_N$. Then, since the intertwining relation above implies that $H_N=S_\Psi^{-1}\,H_N^\dagger\,S_\Psi$, we get
$$
\left<f,H_N\, g\right>_\Psi=\left<S_\Psi^{1/2}\,f,S_\Psi^{1/2}\,H_N\,g\right>=\left<S_\Psi\, S_\Psi^{-1}H_N^\dagger\,S_\Psi\,f,g\right>=$$
$$=\left<S_\Psi^{1/2}\,H_N\,f,S_\Psi^{1/2}\,g\right>=\left<H_N\,f, g\right>_\Psi,
$$
for all $f,g\in\Hil_N$. Hence $H_N$ turns out to be self-adjoint with respect to $\left<.,.\right>_\Psi$.

Alternatively (and more straightforwardly), formula (\ref{54}) shows that the effective hamiltanians considered here and producing damping are
similar (but not unitarily equivalent) to self adjoint operators. A natural and quite interesting question to consider is therefore: is this a
general requirement? We have no answer now, but this is part of our works in progress.

\section{Conclusions}

After some general remarks on the time evolution  driven by a non self-adjoint hamiltonian, we have shown how this can be used in the analysis
of the decay of finite dimensional systems in connection with pseudo-fermions. In particular, in the examples discussed here, we have seen that
the effective hamiltonians describing damping are related, in a very direct way, to pseudo-hermitian operators. We have also shown that the
Heisenberg representation can be conveniently adopted in the analysis of the time evolutions of these systems.

The next natural step of our research will be to check how much of this approach can be extended to infinite dimensional spaces and to try to
answer to the following, rather general, question: can any effective damping be described in terms of some pseudo-hermitian hamiltonian?

\section*{Acknowledgements}

The author acknowledges financial support by the MIUR.

\end{document}